# Exchange bias in molecule/Fe$_3$GeTe$_2$ van der Waals heterostructures via spinterface effects

*Junhyeon Jo\*, Francesco Calavalle, Beatriz Martín-García, Fèlix Casanova,*

*Andrey Chuvilin, Luis E. Hueso\*, and Marco Gobbi\**

Dr. J. Jo, F. Calavalle, Dr. B. Martín-García, Prof. F. Casanova, Prof. A. Chuvilin, Prof. L. E. Hueso, Dr. M. Gobbi

CIC nanoGUNE, 20018 Donostia-San Sebastian, Basque Country, Spain

E-mail: j.jo@nanogune.eu; l.hueso@nanogune.eu; m.gobbi@nanogune.eu

Prof. F. Casanova, Prof. A. Chuvilin, Prof. L. E. Hueso, Dr. M. Gobbi

IKERBASQUE, Basque Foundation for Science, 48013 Bilbao, Basque Country, Spain

Dr. M. Gobbi

Centro de Física de Materiales (CFM-MPC) Centro Mixto CSIC-UPV/EHU, San Sebastián/Donostia 20018, Spain

Keywords: 2D magnets, Fe$_3$GeTe$_2$, spinterface, exchange bias, hybrid van der Waals heterostructures, Co-phthalocyanine (CoPc)

**Abstract**

The exfoliation of layered magnetic materials generates atomically thin flakes characterized by an ultrahigh surface sensitivity, which makes their magnetic properties tunable via external stimuli, such as electrostatic gating and proximity effects. Another powerful approach to tailor magnetic materials is molecular functionalization, which leads to hybrid interface states with peculiar magnetic properties, called spinterfaces. However, spinterface effects have not yet been explored on layered magnetic materials.

Here, we demonstrate the emergence of spinterface effects at the interface between flakes of the prototypical layered magnetic metal $Fe_3GeTe_2$ and thin films of paramagnetic Co-phthalocyanine. Magnetotransport measurements show that the molecular layer induces a magnetic exchange bias in $Fe_3GeTe_2$, indicating that the unpaired spins in Co-phthalocyanine develop antiferromagnetic ordering by proximity and pin the magnetization reversal of $Fe_3GeTe_2$. The effect is strongest for a $Fe_3GeTe_2$ thickness of 20 nm, for which the exchange bias field reaches -840 Oe and is measurable up to approximately 110 K. This value compares very favorably with previous exchange bias fields reported for $Fe_3GeTe_2$ in all-inorganic van der Waals heterostructures, demonstrating the potential of molecular functionalization to tailor the magnetism of van der Waals layered materials.

## 1. Introduction

The exfoliation of layered magnetic materials yields ultrathin single crystalline flakes,[1,2] which represent an ideal playground to explore magnetism in reduced dimensions.[3–6] In this materials family, $Fe_3GeTe_2$ (FGT) is a particularly intriguing compound, since it is a ferromagnetic metal characterized by a Curie temperature close to the room temperature and strong out-of-plane anisotropy.[7–10] Moreover, the magnetic properties of FGT are tunable, as they can be modified either electrically, applying a gate voltage [11] and a large electrical current,[12] or through magnetic proximity effects.[13–16] In particular, when interfaced with antiferromagnetic layered materials, FGT displays an increased coercivity and exchange bias, [13–16] which are the prototypical manifestation of magnetic proximity[17–19] and are key elements in spintronic devices.[20]

Another attractive approach to tune the properties of a magnetic surface is molecular functionalization.[21,22] The interface between magnetic materials and molecules host hybrid states characterized by a peculiar magnetic texture, named spinterface,[21,22] which leads to radical changes on the magnetic properties of both the molecular layer[23–30] and the magnetic material.[27–33] So far, the ferromagnetic layers used for investigating spinterface effects are typically films of $3d$ metals or oxides with dangling bonds on the surface, which result in non-ideal interfaces with molecules. Moreover, until recently, the molecular side of a spinterface has been the main target of research due to its easily tunable electronic properties,[23–29] whereas the possibility of tailoring the magnetism of ferromagnetic materials has yet to be fully exploited.

Layered magnetic materials are excellent candidates for developing a spinterface in view of their tunable magnetism and their single crystalline nature, which offer the possibility to form highly controllable interfaces with molecules[34,35] via the so-called van der Waals epitaxy.[36–39] Indeed, hybrid heterostructures based on atomically sharp 2D material/molecule interfaces have been widely used to tailor the opto-electronic and transport properties of non-magnetic layered materials.[40–45] However, so far the possibility to tune the properties of a layered magnetic material through the magnetic interactions at a van der Waals spinterface has not yet been experimentally demonstrated.

Here, we report on the emergence of spinterface effects between molecular films of paramagnetic Co-phthalocyanine (CoPc) and a few-nm-thick FGT flakes. The molecular layer induces a negative magnetic exchange bias in FGT, indicating that the paramagnetic spins in CoPc develop antiferromagnetic interlayer ordering by proximity, and couple ferromagnetically to FGT. This spinterface effect, which is detected via magnetotransport measurements, is characterized by an exchange field as large as -840 Oe, and a blocking temperature of 110 K. These characteristic values are both among the largest reported in the context of layered magnetic materials.[13–16,46,47] The creation of a molecular spinterface with a layered magnetic material opens a new avenue to tailor the magnetism of layered materials towards hybrid low-dimensional functional devices.

## 2. Results and Discussion

Several studies show that exchange bias emerges at the interface between ferromagnetic metals and organometallic molecules[27–29,48], due to a proximity-induced antiferromagnetic ordering of molecular spins, which pin the magnetization reversal. Among the organometallic compounds used for these exploring spin-effects[49], metallo-phthalocyanine (MPc) are a versatile class of planar semiconducting molecules widely utilized in research and industry (**Figure 1**a). The metal ion ($M^{2+}$) at the center of the molecule, usually a transition metal such as Co, Cu, or Zn, provides diverse energy levels, charge mobility, and spin states.[49–51] To explore the emergence of exchange bias at a FGT spinterface (Figure 1b), we used CoPc, which is a paramagnetic molecule with a spin of $S = ½$. Importantly, it was shown that the CoPc spins are prone to develop antiferromagnetic ordering up to a relatively high temperature ($J/k_B \sim 107$ K)[52,53].

For the fabrication of devices based on CoPc/FGT interfaces, we first deposited CoPc molecules on a Si/SiO$_2$ substrate with Au electrodes prepatterned in a Hall bar design (see Methods for details). Viscoelastic stamping with polydimethylsiloxane (PDMS) was used to mechanically exfoliate a FGT flake and transfer it onto the CoPc-covered substrate. The so-obtained CoPc/FGT structure was encapsulated with a hBN flake. The stamping processes were performed in an Ar-filled glovebox, to prevent the oxidation of FGT in air. **Figure 2**a shows the optical image of a fabricated CoPc/FGT device. The 6-nm-thick CoPc layer covered the whole area observed in this image, while the 20-nm-thick FGT and hBN layers are highlighted by a red and black line, respectively.

The integrity of the CoPc film was confirmed through microRaman spectroscopy. Figure 2b shows a comparison of the Raman spectra measured in different regions of the CoPc film, either covered by hBN (CoPc/hBN) or by FGT and hBN (CoPc/FGT/hBN). In both regions, the spectra display the typical features of MPc molecules. In particular, the peaks of 1466 cm$^{-1}$ and 1542 cm$^{-1}$ represent the $B_{1g}$ and $B_{2g}$ mode of the CoPc molecule, and correspond to the C-N stretching directly associated to the central Co ion (detailed description in Figure S1, Supporting Information).[54] These Raman features were also observed in an uncovered thin film of CoPc molecules, indicating that the stamping process did not significantly affect the integrity of the molecular layer (Figure S1, Supporting Information).

Figure 2c displays the Raman features of a hBN-capped FGT flake transferred on a CoPc film or on a bare SiO$_2$ substrate. The two flakes displayed the two dominant Raman peaks of FGT at 120 cm$^{-1}$ and 155 cm$^{-1}$ associated to the $A_{1g}$ and $E_{2g}$ vibrations.[55] Even in this case, there was no significant change in the spectra for FGT in contact with CoPc or SiO$_2$. In addition, we mapped the intensity of the CoPc peak at 1542 cm$^{-1}$ in the area denoted by a white dotted line in Figure 2a. The homogeneity of the intensity in the map implies that the CoPc molecules are uniformly distributed in the CoPc/hBN and CoPc/FGT/hBN regions. We note that the lower intensity of the CoPc features in the CoPc/FGT/hBN region is caused by screening from the opaque FGT flake. These data show that the stamping process also maintains the uniformity of the molecular layer.

Scanning transmission electron microscopy (STEM) was employed to gain additional insights on the CoPc/FGT interface. Figure 2e displays a cross-sectional image of a CoPc/FGT/hBN heterostructure obtained by STEM. A low-magnification image provides a picture of the CoPc/FGT/hBN heterostructure on a Si/SiO$_2$ substrate, showing that the CoPc molecules constitute a compact film on SiO$_2$ and form a homogeneous interface with the FGT flake stamped on the top. Figure 2e also shows how the FGT flake is slightly bent to accommodate the height difference in the sample, being in contact with both the top of the pre-patterned electrodes and with the surface of the CoPc layer on SiO$_2$. Moreover, it is clear that the CoPc layer does not evenly cover the rough Au electrode, which in some regions is in direct contact with the FGT flake, ensuring an efficient charge injection. Enlargement of the STEM image in the right section of Figure 2e highlights the formation of a flat and uniform CoPc/FGT layer structure, which is a crucial factor to generate a spinterface effect between these layers.

To investigate the molecular spinterface effect in CoPc/FGT heterostructures, we have characterized the magnetic hysteresis of FGT flakes through magnetotransport measurement, by recording their anomalous Hall effect (AHE). We employed a specific procedure to identify the magnetic interaction at the CoPc/FGT interface. First, we performed a magnetic field-cooling (FC) process to a molecule/ferromagnetic Hall bar device from 300 K to 10 K under an out-of-plane magnetic field ($H_z$) of ±10 kOe, which was used to align the spins in this system. Then, we applied an electrical current ($I$) along a certain physical direction in the FGT device (hereby $x$-axis) and a magnetic field in the out-of-plane direction ($z$-axis) (**Figure 3**a). By measuring the transverse Hall voltage ($V_{xy}$), we were able to record the Hall resistance ($R_{xy} = V_{xy}/I$) as a function of the magnetic field. In this way, we could detect the hysteresis of FGT *via* the AHE and relate it to the magnetic response at the CoPc/FGT spinterface. We notice here that the temperature dependence of the longitudinal resistance ($R_{xx}$) shows a metallic trend typical of FGT (Figure S2, Supporting Information).[7]

Figure 3b shows the magnetic hysteresis of a CoPc(6)/FGT(20) (thickness in nm) heterostructure measured at 10 K after FC with +10 kOe and –10 kOe, respectively. The measured hysteresis loops display a significant shift from the center ($H = 0$) and, in particular, a large exchange bias field $H_{EB} = -840$ Oe at 10 K is recorded after FC with +10 kOe (black line in Figure 3a). Here, the exchange bias field is defined as $H_{EB} = (H_{C+} + H_{C-})/2$, where $H_{C+}$ and $H_{C-}$ indicate the positive and negative coercive field, respectively. When we performed FC with –10 kOe and subsequently measured a Hall resistance at 10 K, the hysteresis loop shifted towards a positive field direction (red line in Figure 3a), for the same amount as in the case of the positive FC. This negative exchange bias, where the direction of a hysteresis loop shift is opposite to the direction of FC, is typically found when a ferromagnetic interfacial coupling develops at the interface between a ferromagnet and an antiferromagnet.[19]

We conducted two supplementary measurements to ensure that the measured exchange bias arises from the magnetic interaction between the layered magnetic materials and the molecular film. Firstly, we fabricated a control sample, FGT(20)/hBN without CoPc molecules, and performed the same FC procedure and AHE measurement. This control sample clearly showed a symmetric hysteresis loop without any exchange bias (Figure S3, Supporting Information). Secondly, although our CoPc/FGT structures were encapsulated by hBN in a Ar-filled glovebox, we could not ignore the possibility of oxidation during transferring the sample to the measurement chamber, which could induce the formation of an oxidized FGT layer and exchange bias.[46] To exclude this effect, we tested a FGT(25) flake intentionally exposed to air 15 minutes without any encapsulation film. In this case, we could not detect any exchange bias nor any noticeable coercive field enhancement (Figure S4, Supporting Information).

The origin of the negative exchange bias at a CoPc/FGT interface can be understood considering that (i) the spins in the CoPc layer acquire long-range antiferromagnetic ordering and (ii) the spins of the CoPc molecules in contact to the FGT surface couple ferromagnetically to it, causing a negative shift in the hysteresis loop. This situation, which is similar to the scenario reported for other organometallic compounds on conventional magnetic materials,[27–29,48] is further validated by analyzing the temperature dependence of the exchange bias.

Figure 3c shows the AHE results measured at different temperatures from 10 K to 200 K after FC with +10 kOe. As the measurement temperature increases, the asymmetry of hysteresis loops decreases and became negligible above 80 K. From the measured loops, we collected the coercive fields and plotted them in Figure 3d and Figure S5 (Supporting Information). Compared to a control FGT device, positive coercive fields in a CoPc/FGT structure deviate significantly from the conventional exponential behavior (solid line in Figure 3d) as temperature lowers. Conversely, negative coercive fields follow well the exponential behavior (Figure S5, Supporting Information). The deviation for positive coercive fields from the exponential behavior as well as the different trend of positive and negative coercive fields confirm the presence of exchange bias. Figure 3e shows that exchange bias fields at different temperatures, collected from the data in Figure 3c, follow the exponential relation $H_{EB} = H_0 \cdot exp(-T/T_1)$, where $H_0$ is the extrapolated value to zero temperature and $T_1$ is a constant. Using this fitting, the blocking temperature (i.e. the starting point of $H_{EB} = 0$) of a CoPc(6)/FGT(20) structure is estimated as 110 K. This value is analogous to the critical temperature in other molecular exchange bias systems based on CoPc[48] and it also corresponds to the estimated exchange energy for the antiferromagnetic ordering of CoPc molecular layers.[53] This agreement with previous results indicates that the exchange bias in FGT indeed arises from the same physical mechanism, i.e. antiferromagnetic ordering in CoPc and its interfacial coupling to FGT.

The thickness of both the ferromagnetic and the antiferromagnetic layers are critical factors determining the exchange bias. **Figure 4**a displays the hysteresis loops measured at 10 K after FC with +10 kOe in different CoPc(6)/FGT($t$) heterostructures, in which the CoPc thickness was kept constant (6 nm) and the FGT one was varied from $t = 10$ nm to $t = 80$ nm. Here, to focus on a coercive field change, we plot the transverse resistance $R_{xy}$ normalized to its value at saturation. First, we note that the coercive field increases for thinner FGT flakes. More interestingly, Figure 4b shows the dependence of the exchange bias on the FGT thickness. For thin flakes ($t < 20$ nm), exchange bias fields increased from $H_{EB} = -62.5$ Oe at 10 nm to a maximum $H_{EB} = -840$ Oe at 20 nm of FGT. Above 20 nm, the exchange bias field decreased gradually down to a value of $H_{EB} = -145$ Oe for 80 nm of FGT. In conventional exchange bias systems, the magnitude of the exchange bias is inversely proportional to the thickness of the ferromagnet as $H_{EB} \sim 1/t_{FM}$.[19] Our CoPc(6)/FGT($t$) system follows this relation in the FGT thickness range from 20 nm to 80 nm (red line in the inset of Figure 4b). The divergence for FGT thicknesses below 20 nm could arise from the lower volume magnetization of FGT compared to metallic ferromagnets such as Co and Fe,[7,46] which might be insufficient to fully activate the antiferromagnetic ordering in the interfacial CoPc layers. Regardless of the thickness of FGT, all the structures exhibited negative exchange bias which highlights the favorable ferromagnetic coupling at the interface between the FGT and CoPc layers.

After having inspected the dependence of the exchange bias on the FGT thickness, we explored its change as a function of the CoPc thickness. **Figure 5**a displays the hysteresis loops measured at 10 K and after FC with +10 kOe in different CoPc($t$)/FGT(20) heterostructures, composed of a FGT flake of 20 nm and different thicknesses of the CoPc layer ($t$ = 2, 4, and 6 nm). In the same panel, we also show the hysteresis of a FGT flake directly transferred on a SiO$_2$ substrate without CoPc (corresponding to $t$ = 0 nm), which displays a coercive field of approximately 3 kOe without exchange bias. For the FGT flake on the 2-nm-thick CoPc layer, we observed a finite exchange bias and a larger coercive field. A similar increase in coercive field is often observed for ferromagnet/antiferromagnet structures characterized by inhomogeneous magnetic textures at the interface.[56,57] Following the same tendency, an even larger coercive field was recorded for a FGT flake on the 4-nm-thick CoPc layer, which was also characterized by a larger exchange bias field. Finally, we found a maximum exchange bias field and a reduced coercive field for the CoPc thickness of 6 nm. We understand this finding considering that the 6-nm-thick CoPc film sustains antiferromagnetic ordering and forms a continuous interface with FGT, so it pins more efficiently the FGT layer. Figure 5b displays the increase in an exchange bias field with a CoPc thickness from $H_{EB}$ = 0 Oe (without CoPc) to $H_{EB}$ = –840 Oe (at CoPc 6 nm).

Furthermore, the quality of a magnetic heterointerface at a ferromagnet/antiferromagnet structure can be evaluated by the ratio between the exchange bias and the coercive field ($H_{EB}/H_C$).[56,57] Therefore, Figure 5b also shows the $H_{EB}/H_C$ ratio in the CoPc($t$)/FGT(20) structures. The $H_{EB}/H_C$ ratio reaches a remarkably large value of 0.34, corresponding to the CoPc thickness of 6 nm. This value is the highest among the reported exchange bias structures based on FGT (see Table S1 for comparison with other systems),[13–16,46,47] indicating the high quality of the magnetic CoPc/FGT spinterface.

## 3. Conclusion

In this study, we demonstrated the emergence of spinterface effects in hybrid van der Waals heterostructures composed of a CoPc film interfaced with a flake of a layered ferromagnet FGT. The formation of a homogeneous CoPc layer and of a flat and sharp CoPc/FGT interface, confirmed through micro-Raman spectroscopy and STEM, are ideal for the emergence of magnetic interaction between the unpaired spins in CoPc and the ferromagnetic FGT. Magnetotransport measurements performed following a FC procedure indicate that the interfacial magnetic interaction induces exchange bias in FGT persisting up to 110 K. This blocking temperature is analogous to the reported exchange energy of antiferromagnetic ordering of CoPc layers,[48,53] suggesting that the CoPc layer develops antiferromagnetic ordering which couples ferromagnetically to the FGT surface, pinning its magnetization. This is corroborated by analyzing the AHE of CoPc/FGT heterostructures with different CoPc thicknesses, which evolve from an increase in coercive field for thin CoPc layers to a fully developed exchange bias for thicker layers. The $H_{EB} = -840$ Oe, recorded for a 20-nm-thick FGT flake, is the strongest value reported so far for layered magnetic materials, highlighting the superior quality of the CoPc/FGT spinterface. Our results show that hybrid van der Waals heterostructures composed of layered magnetic material interfaced to organic molecules represent an ideal materials platform to develop high quality spinterface effects, which would be a key element in atomically precise multifunctional structures for practical device application.

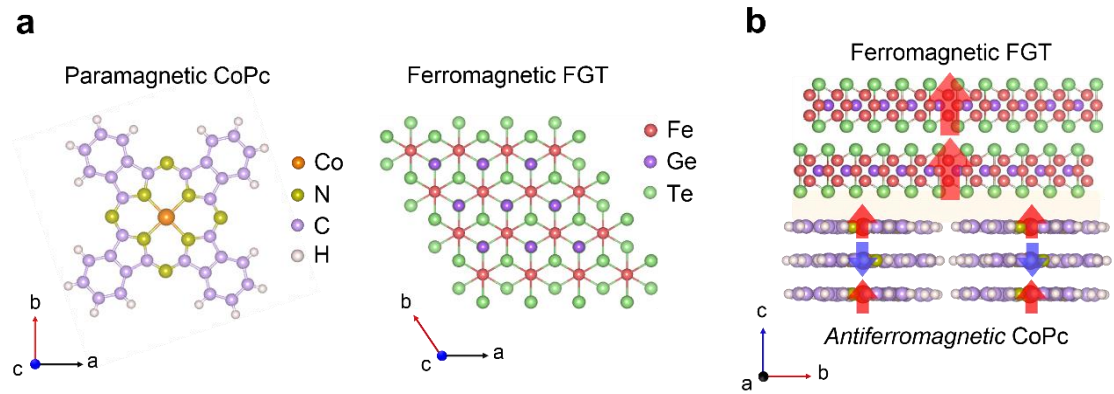

**Figure 1**. Schematics of a hybrid van der Waals heterostructure resulting from interfacing CoPc and FGT. a) Chemical structures of a planar and paramagnetic molecule CoPc and of a layered ferromagnet FGT (top view). b) Schematic rendering of a hybrid CoPc/FGT heterostructure (lateral view). The emerging spinterface mechanism is also sketched: the ferromagnetic FGT flake (red arrows in the picture) interfaced with the CoPc layers activate their antiferromagnetic molecular ordering (red/blue arrows), which in turn pins the FGT magnetization introducing exchange bias.

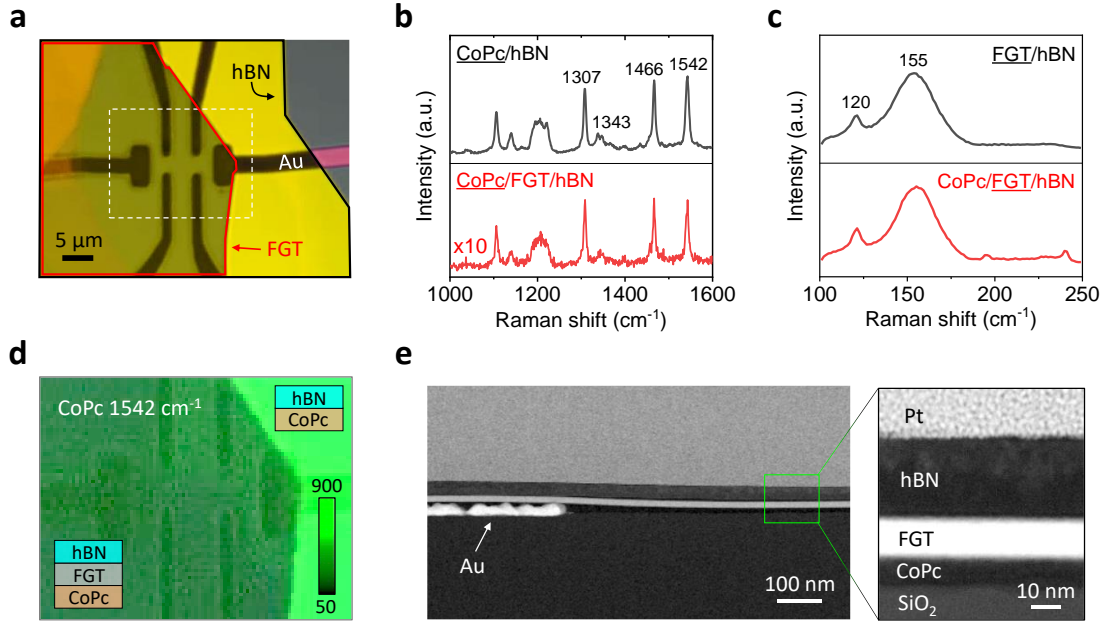

**Figure 2**. Structural characterization of a hybrid CoPc/FGT heterostructure. a) Optical image of a CoPc/FGT structure encapsulated by hBN. The CoPc layers cover all the substrate area in this image, and the red and black lines indicate the FGT and hBN flakes, respectively. b) Representative Raman spectra for CoPc with and without FGT layers. Raman peaks representing intra-molecular vibrations (1300 cm$^{-1}$ to 1600 cm$^{-1}$) are similar in both the CoPc/hBN and CoPc/FGT/hBN regions of the sample. c) Representative Raman spectra for FGT with and without CoPc layers. The typical FGT peaks at 120 cm$^{-1}$ and 155 cm$^{-1}$ appear unaltered in both FGT/hBN and CoPc/FGT/hBN structures. d) Raman map image for CoPc (1542 cm$^{-1}$) in the white dotted line region of Figure 2a, displaying the formation of a homogeneous CoPc molecular layer. The region covered by FGT shows a lower Raman intensity due to the light absorbance from the FGT flke. e) Cross-sectional STEM images of a CoPc/FGT/hBN structure. The interface is flat and homogeneous.

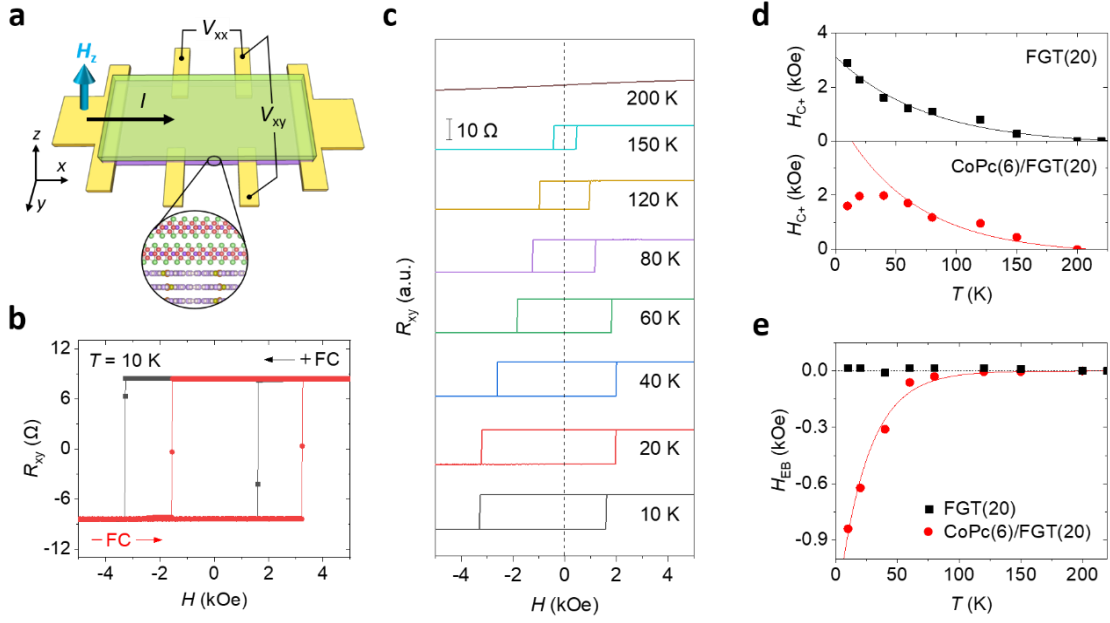

**Figure 3**. Exchange bias in a CoPc(6)/FGT(20) heterostructure. The numbers between parenthesis indicate the thickness in nanometers. a) Sketch of a CoPc/FGT device. An electrical current ($I$) is applied along the longitudinal direction ($x$) and the Hall voltage ($V_{xy}$) is measured transversally (along $y$) while applying an out-of-plane magnetic field (along $z$). b) Hall resistance as a function of the magnetic field at 10 K after FC with +10 kOe (+FC) and –10 kOe (–FC). The hysteresis associated to the AHE of the FGT, is shifted to the opposite directions with respect to the polarity of the FC, and the magnitude of these shifts is identical. c) Temperature-dependent AHE after FC with +10 kOe. The asymmetry of a hysteresis loop decreases as temperature increases. d) Temperature-dependence of a positive coercive field, $H_{C+}$, after FC with +10 kOe in a FGT(20) and CoPc(6)/FGT(20) structure. The control sample of FGT displays an exponential increase of $H_{C+}$ when lowering temperature, while the CoPc/FGT heterostructure displays a deviation from the exponential trend at low temperatures, where the exchange bias emerges. e) Exchange bias fields, $H_{EB}$, as a function of temperature in a CoPc(6)/FGT(20) and in a FGT(20) structure. The exchange bias in the CoPc/FGT follows an exponential behavior with an estimated blocking temperature of 110 K. No exchange bias is recorded in the control sample of FGT.

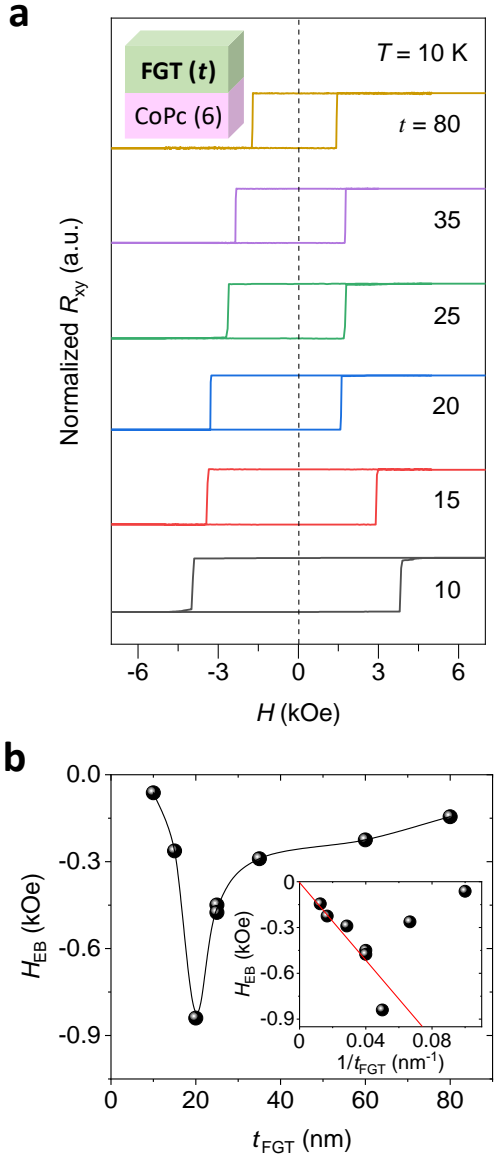

**Figure 4**. Dependence of exchange bias on the FGT thickness. a) AHE measured in different CoPc(6)/FGT($t$) heterostructures composed of a 6-nm-thick CoPc film and a FGT flake with varying thickness (between 10 and 80 nm). b) Dependence of the exchange bias field in the CoPc(6)/FGT($t$) heterostructures as a function of the FGT thickness $t_{FGT}$, as extracted from the curves in (a). Inset: the same data plotted as a function of $1/t_{FGT}$ highlights the proportional relation between $H_{EB}$ with the inverse of the FGT thickness from 20 nm to 80 nm. These data were measured at 10 K after a FC procedure with +10 kOe.

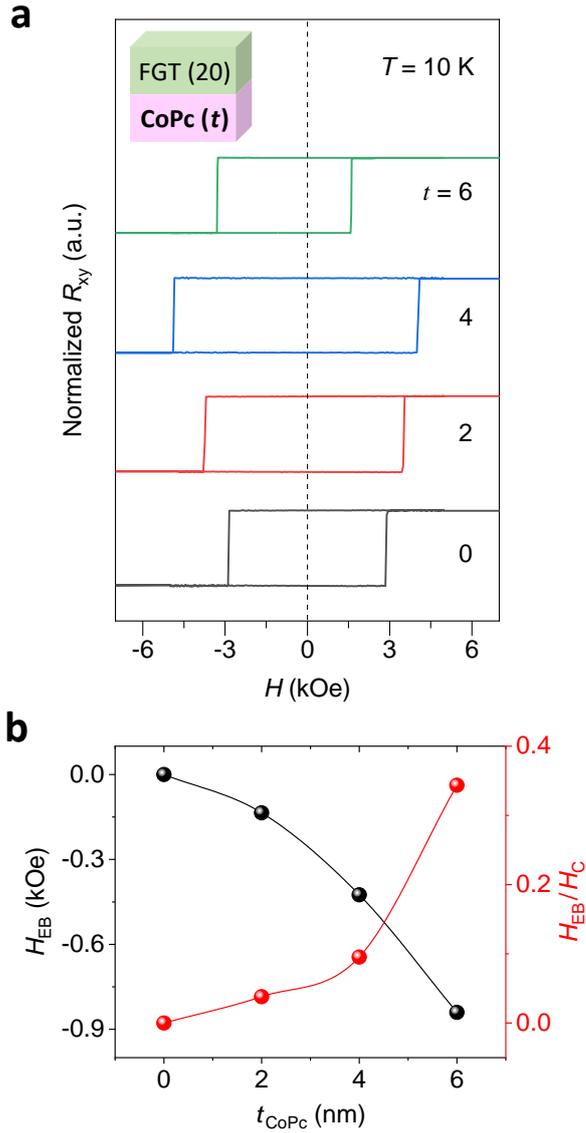

**Figure 5**. Dependence of exchange bias on the CoPc thickness. a) AHE measured in different CoPc(*t*)/FGT(20) heterostructures composed of a 20-nm-thick FGT flake and a CoPc film of varying thickness (between 0 and 6 nm, where 0 nm corresponds to the absence of CoPc. b) Dependence of $H_{EB}$ in CoPc(*t*)/FGT(20) heterostructures as a function of the CoPc thickness $t_{CoPc}$, as extracted from the curves in (a). The $H_{EB}/H_C$ ratio is also shown, where $H_C$ is a coercive field in the measured hysteresis, recorded as a half of the loop width.

**Experimental Section/Methods**

*Sample preparation*: Ti/Au electrodes (2+10 nm) with a Hall-bar geometry were prepatterned on a Si/SiO$_2$ substrate. The channel dimension in the Hall pattern is 10 μm x 2 μm. CoPc molecules (purchased from Sigma Aldrich) were deposited on the electrode-prepatterned Si/SiO$_2$ substrates, previously cleaned with acetone, isopropyl alcohol, and plasma treatment. The deposition rate of CoPc was controlled to 0.1 Å/s while the substrate was maintained at room temperature, and the base pressure of vacuum was 1.2 x 10$^{-8}$ mbar. A FGT flake (HQ graphene) was prepared by mechanical exfoliation using a blue tape (Nitto® SPV224) and transferred to a PDMS film (Gel-pak). The thickness of an exfoliated FGT flake on a PDMS film was characterized by optical contrast in an optical microscope, and confirmed again by using atomic force microscopy (Agilent 5500 SPM). The selected FGT flake was transferred onto the electrode-prepatterned substrate and encapsulated by a hBN flake as the same way. Exfoliation and stamping were performed in an Ar-filled glovebox (H$_2$O and O$_2$ < 0.1 ppm).

*Raman characterization*: Micro-Raman measurements (Renishaw InVia Qontor micro-Raman instrument) were carried out at room temperature using a 100× objective with a sample kept in high vacuum in a Linkam chamber. 532 nm and 633 nm laser excitation sources were used for FGT and CoPc, respectively. A step (*x,y*) size of 0.2 μm x 0.2 μm was used for micro-Raman mapping.

*TEM analysis*: TEM, STEM data was acquired using TitanG2 60-300 electron microscope (FEI, Netherlands) equipped with xFEG, monochromator, image aberration corrector, HAADF STEM detector and Quantum GIF (Gatan, UK). Images were obtained at 300kV accelerating voltage. Cross-section samples of the devices have been prepared by a standard FIB protocol using Helios 600 FIB/SEM (FEI, Netherlands).

*Magneto-transport measurement*: Electrical transport measurements were performed in a physical property measurement system (PPMS, Quantum Design) with a rotational sample stage. An electrical current was applied by a Keithley 6221 and a measured voltage was detected by a Keithley 2182 nanovoltmeter. For a FC procedure, the magnetic field of ±10 kOe was applied at 300 K and a sample was cooled down to a low temperature (10 K) while maintaining the magnetic field. The measurement with a magnetic field sweep started from the applied magnetic field (±10 kOe).

**Acknowledgements**

This work was supported by "la Caixa" Foundation (ID 100010434), under the agreement LCF/BQ/PI19/11690017, by the Spanish MICINN under Project PID2019-108153GA-I00, RTI2018-094861-B-100 and under the María de Maeztu Units of Excellence Program (MDM-2016-0618). This work was also supported by the FLAG-ERA grant MULTISPIN, by the Spanish MCIN/AEI with grant number PCI2021-122038-2A. B.M.-G. thanks Gipuzkoa Council (Spain) in the frame of Gipuzkoa Fellows Program. J.J. acknowledges the support of Basic Science Research Program through the National Research Foundation of Korea (NRF) funded by the Ministry of Education (2020R1A6A3A03039086).

Supporting Information

# Exchange bias in molecule/Fe$_3$GeTe$_2$ van der Waals heterostructures via spinterface effects

*Junhyeon Jo\*, Francesco Calavalle, Beatriz Martín-García, Felix Casanova,*

*Andrey Chuvilin, Luis E. Hueso\*, and Marco Gobbi\**

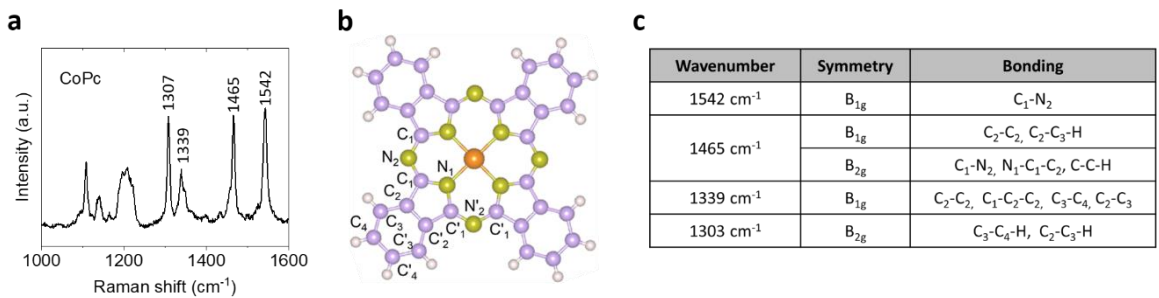

**Figure S1**. Raman characterization of CoPc. a) Raman spectrum of CoPc layers. The features in the range of 1300 cm$^{-1}$ to 1600 cm$^{-1}$ represents intra-molecular vibrations. b) The structure and atomic notation of a CoPc molecule. c) Details about intra-bonding and symmetry for CoPc Raman peaks.[54] All the Raman spectra of the CoPc show identical features to the CoPc in the heterostructure of CoPc/hBN and CoPc/FGT/hBN in Figure 2b.

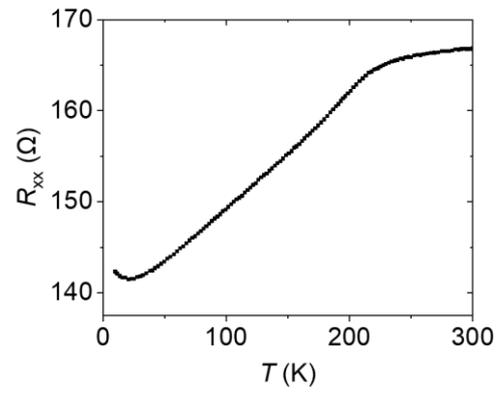

**Figure S2**. Temperature-dependent longitudinal resistance of a CoPc(6)/FGT(20) heterostructure. A kink near 200 K indicates the phase transition of FGT from paramagnetism to ferromagnetism.

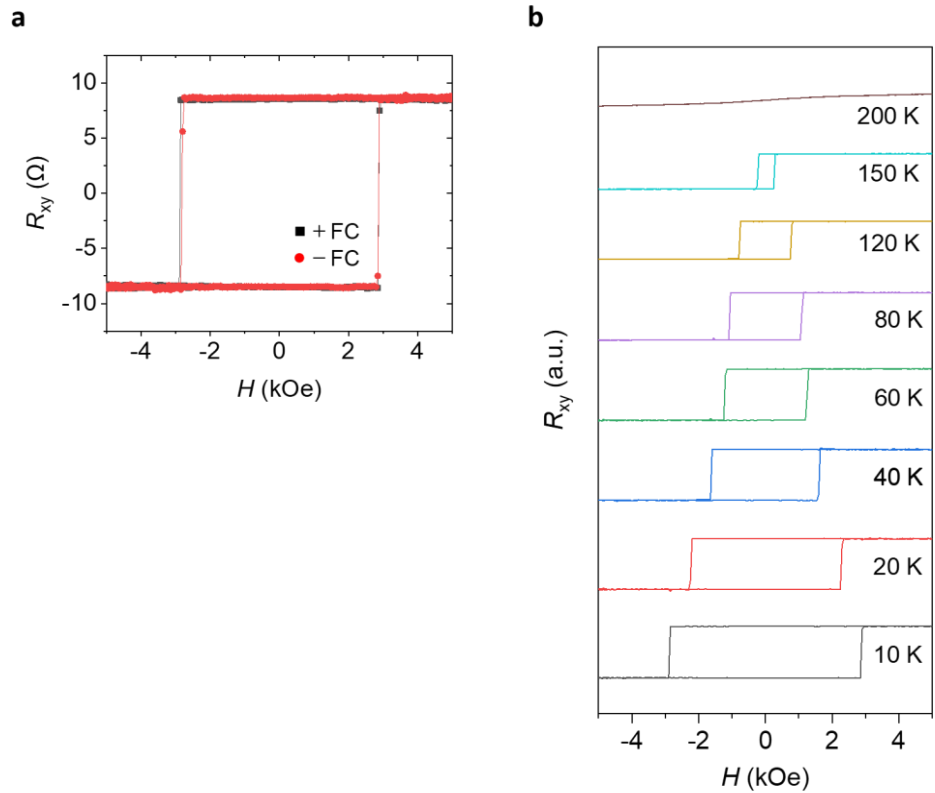

**Figure S3**. AHE measurement on a hBN-capped FGT(20) structure after a FC procedure. a) Hysteresis loop measured at 10 K after FC with ±10 kOe. Without CoPc layers, we found no asymmetry at a positive and negative coercive field. b) AHE measured at different temperatures from 10 K to 200 K after FC with +10 kOe. Exchange bias is not observed in any of the loops.

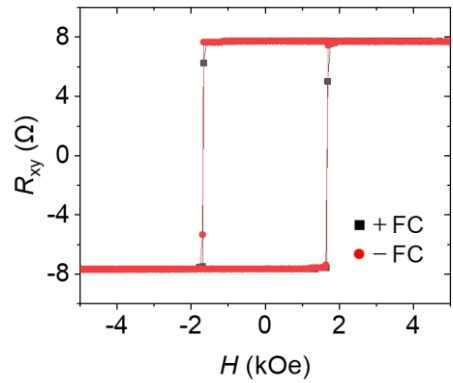

**Figure S4**. AHE measurement on a FGT(25)/O-FGT(*x*) structure at 10 K after FC with ±10 kOe. There is no signature of exchange bias even though an exfoliated 25-nm-thick FGT flake was intentionally exposed to air 15 mins before encapsulated by hBN.

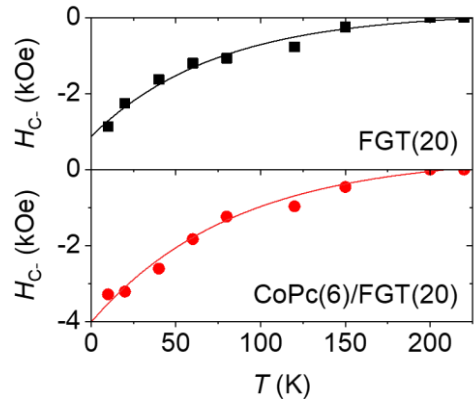

**Figure S5**. Temperature-dependence of a negative coercive field, $H_{C-}$, after FC with +10 kOe in a FGT(20) and CoPc(6)/FGT(20) structure. Both the control sample of FGT and the CoPc/FGT heterostructure display an exponential behavior of $H_{C-}$ on varying temperature.

| Structure (unit : nm) | $H_{EB}/H_C$ | $T$ (K) | AFM type | Reference |
|---|---|---|---|---|
| FGT(30)/CrCl$_3$(15) | 0.2 | 10 | uncompensated | [14] |
| FGT(30)/CrCl$_3$(45) | 0.1 | 10 | uncompensated | [14] |
| FGT(17)/O-FGT(3) | 0.15 | 70 | | [46] |
| FGT(38)/O-FGT(3) | 0.28 | 70 | | [46] |
| FGT(20)/CoPc(6) | 0.34 | 10 | uncompensated | This work |
| FGT(25)/IrMn(2) | 0.04 | 2 | compensated | [47] |
| FGT(23)/MnPSe$_3$(23) | 0.04 | 10 | compensated | [15] |
| FGT(23)/MnPS$_3$(23) | 0.05 | 10 | compensated | [15] |
| FGT(9)/MnPS$_3$(19) | 0.05 | 10 | compensated | [12] |

**Table S1**. $H_{EB}/H_C$ ratio in FGT based exchange bias systems. The ratio is calculated with recorded data in the reference papers.